# AN EXPLICIT TRUST MODEL TOWARDS BETTER SYSTEM SECURITY

Orhio Mark Creado[1], Bala Srinivasan[2], Phu Dung Le[3], and Jefferson Tan[4]

[1]Caulfield School of Information Technology, Monash University, Melbourne, Victoria, Australia
mark.creado@monash.edu

[2]Clayton School of Information Technology, Monash University, Melbourne, Victoria, Australia
srini@monash.edu

[3]Caulfield School of Information Technology, Monash University, Melbourne, Victoria, Australia
phu.dung.le@monash.edu

[4]IBM Research – Australia, Melbourne, Victoria, Australia
jeffetan@au1.ibm.com

## ABSTRACT

*Trust is an absolute necessity for digital communications; but is often viewed as an implicit singular entity. The use of the internet as the primary vehicle for information exchange has made accountability and verifiability of system code almost obsolete. This paper proposes a novel approach towards enforcing system security by requiring the explicit definition of trust for all operating code. By identifying the various classes and levels of trust required within a computing system; trust is defined as a combination of individual characteristics. Trust is then represented as a calculable metric obtained through the collective enforcement of each of these characteristics to varying degrees. System Security is achieved by facilitating trust to be a constantly evolving aspect for each operating code segment capable of getting stronger or weaker over time.*

## 1. INTRODUCTION

Trust is an implicit commodity in the world today. We inherently trust our financial institutions, service providers, and even other motorists without any second thought. However, although synonymous, being trusted and being trustworthy are very different [1]. Trust as a human construct is extremely pliable; but this is not the same case when considering computing systems.

Computing systems execute code that performs operations which produce usable outputs. Each instruction can be considered to be a singular operation. Therefore, a computing system can only be trusted and secure depending on the next instruction it executes. So how can a computing system rely on securing itself from itself?[2].

In this paper we propose a novel concept to alleviate the ambiguity of trust levels associated with executing code so as to ensure better overall system security. To achieve this goal, we primarily define trust in terms of a computing architecture; wherein, trust is no longer defined as a singular attribute but instead as a combination of characteristics which can collectively determine the overall trust level for any operating code. The aim behind this paradigm is to represent trust as an evolving concept within a computing system capable of growing stronger or weaker over time based on past operating performance. In our opinion, this is a significant step away from current models which advocate trust to be a binary outcome based on superficial constructs such as a valid username and password.

The rest of this paper is structured as follows. Section 2 briefly covers some of the relevant literature in the area. Section 3 defines the explicit trust model architecture along with its main component - the trust engine, outlining the various trust categories and trust levels within each category and concludes with a real world application of the proposed model's theory. Section 4 evaluates our proposal with a formal analysis of security vs. performance and provides some real world tangibility by using an asset centric threat model documenting some of the attacks pertinent to system security and how the proposed model aims at resolving them. Section 5 concludes our work and provides directions for future research.

## 2. BACKGROUND

Trust, as a concept, traces its roots back as a psychological and sociological construct. In computing, its definition cannot be applied completely; as machines tend to be programmatic intelligence, the task of quantifying malicious intent becomes more challenging [3]. Bevan [4] proposes that with human-computer interactions there remain many variations of trust and trust levels; but not all of these levels can be accounted for in human-human interactions. Yet, computing systems have aimed to satisfy only a few of these variations at any one time, so how can security be achieved if only partial trust can be achieved?

Trust in computing has been an active area of research for a very long time. One of the most prominent implementations of trusted computing has been the Trusted Computing Group's (TCG) Trusted Platform Module (TPM) [5, 6]; which used a cryptographically secure hardware to perform trusted software operations. A well-known implementation of this hardware platform was observed in Microsoft's Next Generation Secure Computing Base (NGSCB) [7-11]. The drawback of this approach was that it tried to facilitate for a trusted area within an otherwise insecure environment.

Alternative solutions include the implementation of microkernels as proposed by Setapa [12], and Heiser et. al. [13]. A good example of a hybrid approach between hardware and software policy has been proposed in the work of Nie et. al. [14]. The drawback with some of these approaches is that microkernels can be vulnerable during the boot phase of a computing system, and relying on hardware based solutions to implement security models is equivalent to no security if it is possible to compromise the actual hardware [15].

Trust can be defined as a concept with multiple characters [16], the challenge of implementing trust in computing has been the subjectiveness of the term 'trust' in relation to the user. This means that any operating code executed on a system can behave differently at different times depending on the user, the operating environment variables, and the desired outcome being sought. The determination of being trustable is still an open concern with human interactions, so why should computing systems be any different? Real world implementations aren't quite as simplistic, so as to be able to always consistency and accurately reduce the outcome of trust and security to a binary result. It isn't feasible, or possible, to account for all the possible scenarios which must have trusted operations defined. Modern day computing systems and their operations are never static, so why should the definition of trusted operations and trust in computing be?

# 3. EXPLICIT TRUST MODEL

Trust within operations in today's digital age is of paramount importance. Trust enforcing mechanisms with binary outcomes have become a single point of failure leading to the exploitation or compromise of a system. This section elaborates our proposal for explicit trust.

## 3.1 Security as a Combination of Characteristics

The more trusted the executing operations within a computing system, the more secure is the computing system. With this analogy, the explicit trust model defines a set of characteristics, each with its own set of properties, which can collectively determine the trust associated with all operating code within a computing system. It is important to note that although achieving absolute trust is not possible, it is quite possible to achieve near absolute trust through the correct enforcement of each of the identified characteristics and properties. The defined characteristics are as follows:

- Invulnerable

  Invulnerability can be achieved through the reduction in the number of exploitable errors in operating code. This can be practically envisioned through the definition of secure programming languages, through secure coding practices, and through rigorous application testing. Furthermore, all code should be implicitly defined to handle all errors and be responsible for proper allocation and deallocation of resources. Properties include:

  - Defined Bounds: Ensuring that all input parameters comply with expected inputs, errors which exploit programming language vulnerabilities for input data types can be prevented.
  - Handled Exceptions: Ensuring that all output parameters produced comply with expected outputs, errors which exploit programming language vulnerabilities for output handling can be prevented.

- Integrity

  Integrity can be achieved by having accountability standards in place for all operating code. As all usable code serves a specific purpose and has an author; a publicly verifiable metric, such as digital signatures, associated with the operating code should be provided so as to ascertain its ownership and ensure its authenticity to perform its intended purpose. Properties include:

  - Accountability: Ensuring that all operating code must have a valid and publicly verifiable digital signature which can uniquely identify the owner/author of that operating code and can also uniquely identify the integrity of the code.

- Verification

  Being verified can be achieved through rigorous state management by the operating system. Virtualization technology employs similar aspects which facilitate the management of system state. By preventing unauthorized changes in system states, undesirable states of operation arising from unexpected exceptions in operating code can be prevented. Properties include:

  - Managed States: Ensures that all operations executing one instruction at a time do not forget the operating state of the calling instruction/process parent thereby ensuring the correct completion of instructions from start to finish.

- Trustworthy

    Being trustworthy can be achieved through the proper definition of a calculable trust metric associated with any operating code. Initially assigned based on the credibility of its owner and post that, based on historical performance based on correct execution calculable via deterministic trust algorithms. Properties include:
    - Trust Levels: Ensures that all operating code must have a defined trust level which indicates its level of trustworthiness to the system and upon each execution is recalculated and modified accordingly based on the outcome of that execution.

## 3.2 Identifying Trust Categories and Trust Levels

To facilitate for evolving trust, the proposed model also defines a set of trust categories and underlying trust levels associated with each category. The purpose of these categories is to allow the trustworthiness associated with all operating code to either increase or decrease based on historical performance; thereby implicating higher trustworthiness for correct successful operations and lower trustworthiness for incorrect unsuccessful operations. The following trust categories and underlying trust levels have been defined:

- Functional Trust

    This category outlines the basic trust requirements for all operations within a computing system. All operating system code, user application code, and network services code must have a trust level associated with this category. To allow for application scalability, the standardized constructs which outline the fundamental operations for each application can be application specific. Defines the following trust levels:
    - Operational Trust - Is required for system level operations, such as System-system communication and high priority OS operations.
    - Verifiable Trust - Is the basic requirement for all operational code executed by the system or user to be verifiable and accountable.
    - Denied Trust - Is defined for operational components which are not verifiable and accountable; such as malicious operations aimed at exploiting or compromising the computing system, thereby completely preventing their execution on the computing system.

- Transactional Trust

    This category is defined for operational components to constantly evolve their trust levels by serving as an intermediary between two functional trust levels. Trust levels under this category are deterministically calculable based on past historical operations over time. Defines the following trust levels:
    - Transitional Trust - Intermediate between verifiable and operational trust, facilitates evolution of trust for operations with good historical performance.
    - Untrustable Trust - Intermediate between verifiable and denied trust, facilitates evolution of trust for operations with detrimental historical performance. However, to support versatility and scalability, this trust level allows operational components which do not meet all the verification and accountability standards, but without significant operating history to deny execution, to execute within a constrained operating environment.

## 3.3 Defining the Trust Architecture

This section aims to integrate the defined concepts of the proposed security characteristics in conjunction with the proposed trust levels so as to define the explicit trust model's trust architecture. Traditionally computing systems allow for three types of execution modes: System, User, and Guest. Whilst beneficial, these modes do not define any level of granularity between each and often overlap based on the nature of operations. The proposed model advocates the requirement for a trust level to be associated with the operating code rather than the execution mode of the computing system.

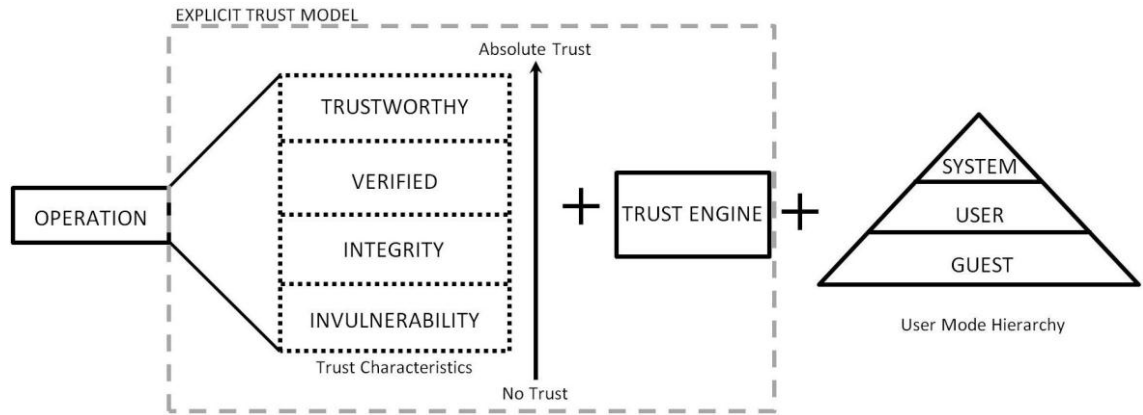

Figure 1. Explicit Trust Model Architecture

Fig. 1 provides a conceptual definition of the proposed trust architecture. The architecture mandates that each operation must be able to satisfy each of the security characteristics by fulfilling their underlying properties. In a realistic scenario these characteristics could be satisfied only to a certain degree and therefore would allow the deterministic calculation of a trust level on a scale from no trust to absolute trust. For this purpose, the architecture defines a trust engine component which acts as an intermediary and facilitates the calculation and determination of the associated trust levels with each operation prior to execution. The last stage of the process is the execution of operating code under one of the execution modes facilitated by the operating system.

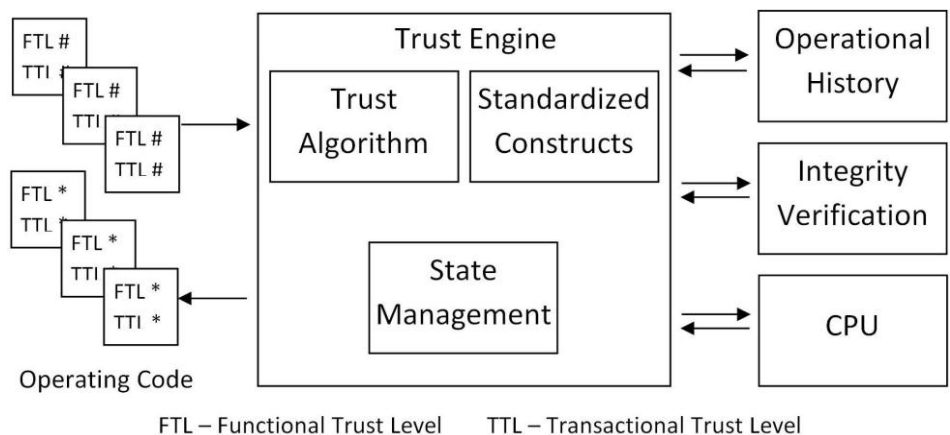

Figure 2. Trust Engine Architecture

The proposed trust engine is the most integral part of the explicit trust model, and Fig. 2 outlines a conceptual definition of the explicit trust model's trust engine architecture. The proposed workings of the trust engine will be defined; its practical implementation, at this stage, is out of scope of this paper.

The trust engine serves as the common link which integrates the various security characteristics with the defined trust levels applicable to operating code. This is facilitated by specifying two calculable trust metrics for each block of operating code; the first being its functional trust level and the second being its transactional trust level. Prior to executing any operating code the trust engine facilitates the following process:

- Verifies its associated functional trust level.
- Verifies its associated integrity signature.
- Determines operating mode and passes instructions for execution.
- On completion of execution, it verifies the state management registry to ensure correct execution.
- Depending on execution outcome, it updates the operation history registry.
- Executes the trust algorithm to deterministically calculate a new transactional trust level based on the operational history.
- Lastly, updates the operating code with a new functional and/or transactional trust level metric.

We now define the various components of the trust engine architecture which facilitate the inner working process of the trust engine. The included components are:

- Operating Code - Defines the basic set of operating instructions which need to be executed on a computing system.
- Trust Algorithm - Deterministic algorithm which takes into account the number of historical executions, correct executions, incorrect executions, owner trust metric, and other key inputs to determine a trust metric for any operating code.
- Standardized Constructs - Defined as an optional customizable add on to the model which would facilitate user or application specific trust requirements.
- Integrity Verification - Verifies the integrity signature of the operating code against the hash of the operating code and the owner's public key.
- Operational History - This registry stores aggregated historical operations for all operating code resident within the computing system serving as input to the trust algorithm's calculation.
- State Management - This registry monitors the execution of processes and forking of parent processes to ensure desirable states of operation and complete execution of instructions from start to finish.
- CPU - Facilitates for the processing and execution of operating code instructions; and accepts required inputs and produces any applicable outputs.

### 3.4 Integrating The Explicit Trust Model

This section aims to further the readers understanding of a possible real world application of the explicit trust model. Fig. 3 illustrates the step by step process of executing operating code based on the associated trust level.

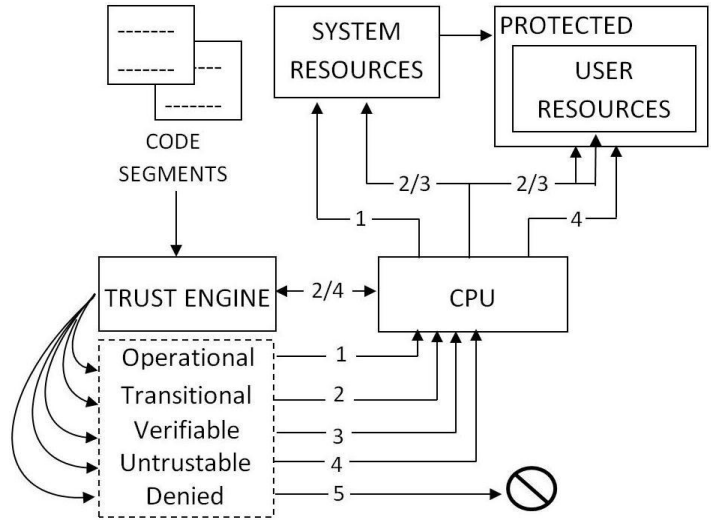

Figure 3. Operational Trust Levels

As observed in Fig. 3, all operations within a computing system can be reduced to operating code making the explicit trust model universally applicable to all operations, users, or even components within a computing system. By associating a trust level with each object it facilitates for a model whereby past operations dictate future access; thereby allowing for an evolving level of trust.

System code is initially owned by its author/owner and once installed on the system changes ownership to the system thereby preventing all future modifications of system level operating code by any user. Updates to system level code would require verification of the original owner in order to allow for modifications of system level code files. By assigning an operational trust level, all operating code within this trust level definition would have access to all system and user level resources.

System and user code which belongs to third parties are always defined with a verifiable trust level provided they have correct integrity signatures. The transactional trust level allows this operating code to evolve to an operational trust level wherein access to system resources might be required in order to perform system level operations. The transactional trust level determines the access to protected user resources and/or system resources.

System and user code which consistently encounter errors or detrimentally affect system state are categorized with a denied trust level wherein all operating code with this trust level is not allowed to be executed on the computing system. To facilitate for operating code without sufficient operating history and/or without verification signatures the untrustable trust level allows for execution of these instructions within a protected environment wherein system level access is completely restricted.

## 4. EVALUATION

In this section we proposal a more formal evaluation of the proposed model for both security and performance. For real world tangibility, we also provide a concise asset centric threat analysis of the model with the emphasis on the computing system. Lastly we conclude the section with an objective discussion of the proposed theory.

## 4.1 Security vs. Performance Analysis

Let us assume that an operating code $C$ is comprised of $n$ lines of operating code, which is a total of $n.k.[O]$ operations; such that

$$C \equiv n \equiv n.k.[O] \tag{1}$$

represents the total number of operations for $n$ lines of code where

$$k = \sum_{i=1}^{n} [O]_{n_i}/n$$

is the coefficient of the number of instructions per line of code.

Eq. (1) therefore represents the default number of instructions to be executed without any additional security enforcing properties.

The explicit trust model calculates security of the model using a probabilistic approach due to the inverse relationship between security and performance. We reduce both metrics to the number of operations being performed in order to deterministically evaluate the additional overhead. Keeping this mind we can represent the following:

$$Perf = n.k.[O] \quad , \quad Sec = 1 - \frac{1}{n.k.[O]} \tag{2}$$

where security is calculated as the probability of finding a single error in any operating code subtracted from the probability of code execution.

Each security enforcing characteristic within the explicit trust model can further be enforced with the addition of additional lines of code to the basic set of operating instructions. We can transform Eq. (1) for each property to represent the total number of additional instructional overhead as follows:

$$C + P_i \equiv \frac{n}{l_i} \equiv \frac{n.k.[O]}{l_i} \tag{3}$$

where $l$ is the additional number of lines of code added to $n$ for property $P_i$

Accounting for each of the defined characteristics within the explicit trust model, we can transform Eq. (2) as follows:

$$Perf = \left\{1 + \frac{1}{l_1} + \frac{1}{l_2} + \frac{1}{l_3} + \frac{1}{l_4}\right\}.n.k.[O] \quad , \quad Sec = 1 - \left[\frac{1}{n.k.[O]}.\{1 + l_1 + l_2 + l_3 + l_4\}\right] \tag{4}$$

where each of the properties of Invulnerability, Integrity, Verification, and Trustworthy have been numerically represented.

Fig. 4 outlines the trade-off between performance and security for the proposed model. The graph depicts the deterministic curve which defines the increase in security with a slight decrease in performance. Since all operating code must be executed in order to be functional, the depicted graph is directly based on the number of operations irrespective of the size of the executing code; thereby facilitating for the evaluation of the additional overhead required in terms of ascertaining additional levels of security for the minimal trade-off in performance.

Furthermore, the evaluation methodology provides for an objective overview of a deterministic vs. probabilistic model; due to the nature of computing systems wherein performance degradation is the direct result of increased operations. However, lapses in security should be based on a probabilistic model; as the mere existence of a vulnerability does not imply exploitation without other key factors being supportive as well.

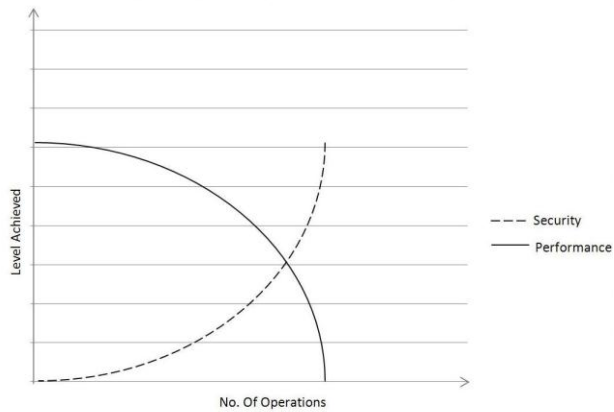

Figure 4. Performance vs. Security Trade-off

Lastly, the proposed work allows for the determination of optimums so as to maintain the balance between security and performance to ensure the usability of a computing system without compromising user friendliness. Furthermore, the abstraction of security enforcing characteristics away from the end user ensures that security does not remain as an optional add-on within a computing system.

## 4.2 Asset Centric Threat Model

In this section we aim to provide the reader with some real world tangibility by proposing the possible feasibility of the proposed model and its application towards preventing real world threats affecting modern day computing systems.

### 4.2.1 Attack Process Flow

Computing systems are processing stations for data - performing operations, and producing desired output or errors. Abstracted within this simplistic view is the attack path used to compromise the system. All attacks must exploit specific inputs so as to compromise a system. Figure 5 graphically outlines the perceived vs. actual process flow of an attack as it happens within a computing system.

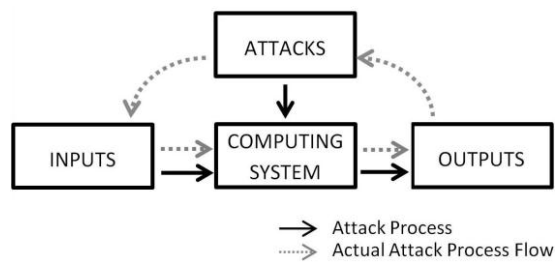

Figure 5. Attacker's Process Flow

### 4.2.2 Threat Identification and Mitigation

With regards to the defined attack process flow, we isolate and outline the various interacting component for the proliferation of trust within a computing system pertinent to our threat model:

- Inputs - Are classified as data needing to be processed by the execution of some code. The issuer of the data or instruction, whether internal or external, is irrelevant to the operation and is therefore not an input.

- Outputs - Post execution of any instruction, the computing system is capable of producing the following outputs: data, errors, or other processes.
- Attacks - The following threats, applicable to system security, have been identified: service disruption, privilege escalation, data theft/manipulation, system corruption, protocol exploitation.

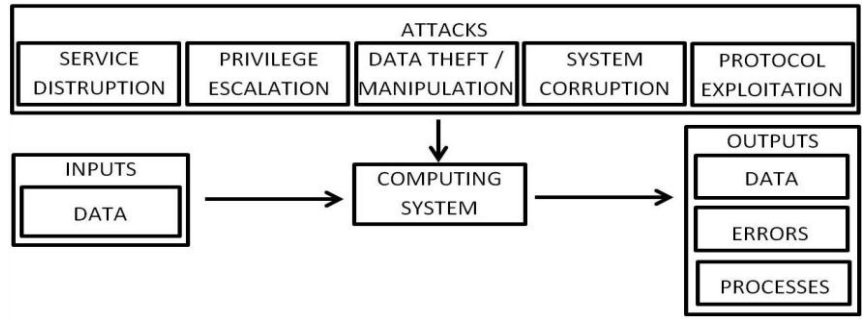

Figure 6. Computing System Threat Model

We identify five main attacks which target computing systems specifically, and represent them in Fig 6 to represent how they relate to our attack process flow. For conciseness we represent the threat mitigation process for these attacks in Fig 7 without taking into account the threat trees for each attack.

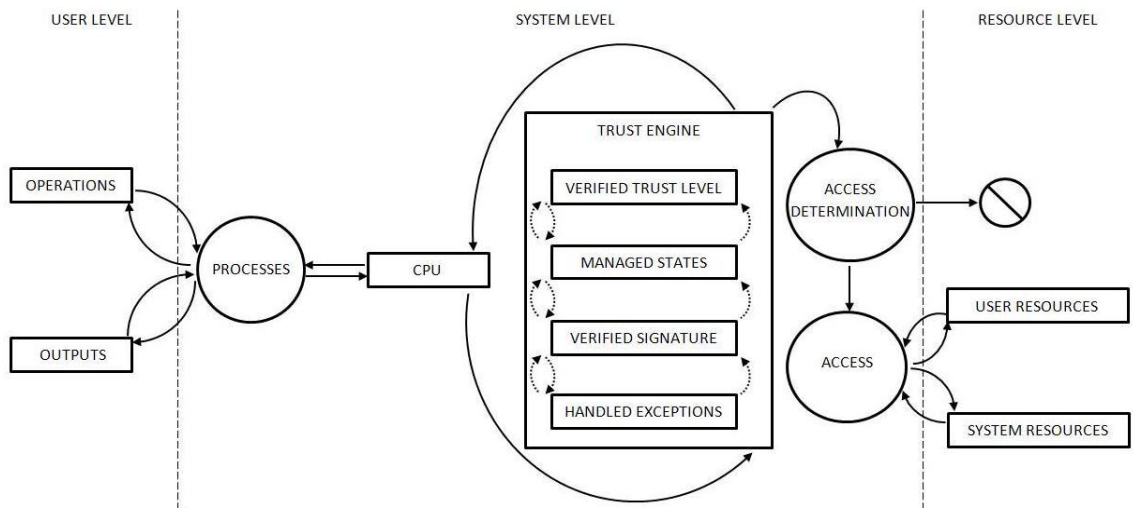

Figure 7. Explicit Trust Threat Mitigation Model

The proposed analysis segments the attack process into three stages: User, System, and Resource. Whilst majority of the actual process might happen at the system level, most attacks target exploiting the resource level by gaining access at the user level. The prevention of these attacks is proposed via the means of the trust engine's trust level determination process which can determine if each of the underlying security enforcing properties is satisfied. By defining a linear progression of characteristics for each operation within the explicit trust model, the trust engine facilitates for a semi-hierarchical approach towards the fulfilment of trust properties to ensure overall system security. The following are broad definitions of these attacks and their mitigations within the proposed model:

- Service Disruption

  Are targeted towards disrupting basic operations; examples include denial-of-service and distributed denial-of-service. The exploitation is targeted towards the communication protocol used and the requirement to acknowledge and respond to all incoming requests. These attacks can be prevented by the integrity and verification properties of the explicit trust model. By validating dual authentication and ensuring state management for all processes these attacks can be circumvented to ensure that all systems communicating with the server can be identified and be held accountable for their actions.

- Privilege Escalation

  Are targeted towards gaining unauthorized access; an example is a buffer overflow attack. The exploitation is targeted towards finding and exploiting coding flaws by passing modified inputs to overwrite memory registers. These attacks can be prevented by the invulnerability, integrity, verification, and trustworthy properties of the explicit trust model. By ensuring that all programming code has proper error handling and resource utilization code in place and by ensuring that all code has an identifiable owner who can be trusted via means of a trust metric associated with the application code. Furthermore, proper state management to ensure instructions finish in order can further prevent these types of attacks.

- Data Theft / Manipulation

  Are targeted towards stealing user data or information; examples include viruses, trojans, spyware/malware etc. The exploitation is targeted towards covert exploitation under the pretence of some other legitimate operation. These attacks can be prevented by the integrity and trustworthy properties of the explicit trust model. Any operating code through covert channels would not be signed with any integrity signature; and furthermore, any default trust metrics associated with these would only be at the verifiable level, which would allow only protected execution thereby thwarting any system level exploitation.

- System Corruption

  Are targeted towards rendering a system unusable; an example is bios corruption. The exploitation is targeted towards overwriting the master boot record thereby rendering the next start-up unable to load. These attacks can be prevented by the integrity and trustworthy properties of the explicit trust model. Any system level changes would require the original author's verification of the operating code. Code affecting the boot load process would ideally be required to have vendor integrity and trustworthy metrics assigned.

- Protocol Exploitation

  Are targeted towards exploiting vulnerabilities in communication protocols; examples include ping of death, certificate forging, session hijacking, scripting etc. The exploitation is targeted towards system modification, disruption, or compromise. These attacks can be prevented by the verification property of the explicit trust model. By ensuring that all processes have a managed state of execution, any variations can be trapped and terminated so as to prevent undesirable states of operations.

## 4.3 Discussion

One of the biggest challenges in theoretical computer science is the evaluation of a proposal so as to ascertain the viability of the idea. Our approach to resolve this has been to provide a different perspective to the reader from a conceptual viewpoint with links to practical applications. Although some of the proposed concepts might seem like existing open challenges in the computing industry the proposed work targets resolving them from a more fundamental point of view which is the underlying source of the vulnerability rather than trying to propose a fix for any specific vulnerabilities. By adopting this approach, our goal remains to propose a model which can be independent of the underlying platform, operating system, application, or component.

For conciseness of this paper, many proposed concepts specifically in the threat model's asset centric approach have been condensed; nonetheless, most of these are implementable via modifying the operating code for most commodity programs and signature verification is currently handled by most operating systems. By reducing our proposal to the most fundamental unit of operating code we allow for the definition of security enforcing characteristics by modifying the existing code. The mammoth task of fixing real world systems is perhaps out of scope of the proposed work; but the argument remains is that if we could fix existing issues we wouldn't still have them. The very fact that vulnerabilities still exist within computing systems points to the fact that the underlying infrastructure might need changing and although perhaps already in the works by big vendors for the not so distant future, this paper has aimed at providing a more conceptually sound, yet practically realisable model to further the state of secure computing.

## 5. CONCLUSION

If the existing paradigms for ensuring trust and security within computing systems were adequate, we wouldn't have as many vulnerabilities and exploitations of systems happening all over the world. Identity theft wouldn't be an issue, man-in-the-middle attacks wouldn't exist, and financial crime would be non-existent. However, that would be an ideal world scenario, but for now there exists a need for our computing infrastructure to evolve to the next level rather than patch existing technology with band-aid solutions which sometimes introduce new vulnerabilities in the process.

In this paper we have proposed a novel approach towards promoting system security by ensuring trusted operations through the proliferation of trust explicitly. We reduce higher order systems to the basic fundamental units of operating code so as to be able to define a linear set of properties which collective define trust as a combination of individual characteristics, rather than viewing it as a singular attribute. Through this approach, we define a process for the deterministic calculation of trust levels based on the degree of satisfaction of each of the properties underlying each of the identified characteristics. By rendering trust as a deterministic metric calculable based on past historical performance, we facilitate for a paradigm of evolving trust within a computing system which can evolve to grow stronger or weaker over time depending on past executions. Furthermore, we evaluate our proposal for the trade-off between security and performance by alleviating the ambiguity between the deterministic vs. probabilistic approach by reducing both aspects to the number of instructions executed we are able to provide a more viable benchmark for comparison which is logically sound.

In our opinion, there remains a large void for secure operations within computing systems with the growing diversity of devices and platforms. Through the incorporation of the proposed model it remains feasible to define security at the core of all operations within a computing system rather than as an add-on aspect dependent on the user. Our plans for continued work in this area include defining a framework for secure computing which is capable of incorporating trust as a fundamental component of its operation. We also have plans to publish our idea of a practical way to realize the proposed model within a computing system. Also in the works include the development of a deterministic trust algorithm which is capable of providing a calculable metric as a trust level using statistical and probabilistic models based on past operational history.